# Efficient Parallel Schemes for Cell-free Massive MIMO Using Radio Stripes

Yihua Ma[1,2], Zhifeng Yuan[1,2], Guanghui Yu[1,2], Yijian Chen[1,2]
[1]ZTE Corporation, Shenzhen, China
[2]State Key Laboratory of Mobile Network and Mobile Multimedia Technology, Shenzhen, China
Email:{yihua.ma, yuan.zhifeng, yu.guanghui, chen.yijian}@zte.com.cn

*Abstract*—Cell-free massive MIMO provides ubiquitous connectivity for multiple users, and implementation using radio stripes is very efficient. This paper proposes a parallel scheme for radio stripes to allow access point to do signal processing simultaneously. Simple maximum ratio (MR) processing works in this scheme, but its performance is bad. Therefore, this paper further proposes two efficient parallel schemes to gain better performance. The first is interference-aware MR processing with a tailored user-centric strategy. The second is distributed regularized zero-forcing (D-RZF) algorithm which derives from LMMSE. Simulation results show that the proposed parallel schemes gain better performance than existing works.

## I. INTRODUCTION

Massive multi-input multi-output (MIMO) is an efficient method to achieve extremely high spectral efficiency, and has been proved to be a very successful 5G technology. It still plays an important role in beyond 5G and 6G [1] to fulfill the key performance indicators of higher spectral efficiency and higher connection density. Many novel schemes using massive MIMO are proposed to further improve the performance. Cell-free massive MIMO [2] is a promising one among them.

Conventional massive MIMO usually deploys a large number of antennas at base station (BS). It is named collocated massive MIMO, which has relatively low deployment cost. However, far-end users suffer from large path loss. To solve this problem, cell-free massive MIMO [2] was proposed. Cell-free massive MIMO is one kind of distributed antenna system where a large number of access points (APs) are deployed at different positions. Each AP can have single or several antennas. In the uplink, AP simply processes the received signal and then delivers it to a central processing unit (CPU). Then, CPU recovers transmit data from different users. Cell-free massive MIMO is able to gain diversity against path loss and provide uniformly good performance for different users.

One classical type of cell-free massive MIMO is that APs connect to CPU via their own front-hauls. In early times, MR processing was advocated for its simplicity [2]. User-centric (UC) strategy [3] can also be applied in MR to reduce the front-haul. Using UC, each user only requires signals from a fraction of APs, and the performance can be even better than original MR in some special cases. [4] compared different levels of cell-free implementations and found that a centralized scheme with optimal linear minimum-mean-square-error (LMMSE) processing achieves both high performance and low front-haul loading. Furthermore, low-resolution analog-to-digital converters (ADCs) was also suggested to reduce both power consumption and cost of cell-free massive MIMO [5] .

Another cell-free implementation method using radio stripes [6] was also proposed for dense scenarios, e.g., stadiums, stations and malls. Multiple APs share one front-haul cable for synchronization, data transmission and power supply. A sequential processing [7] named normalized LMMSE (N-LMMSE) was proposed for radio stripe cell-free scheme, which makes neighboring APs cooperate to gain a better performance. However, AP requires to wait for the results from the previous-stage AP, which brings large processing latency especially when the number of APs is large. This paper focuses on radio stripe cell-free schemes which not only gains a good performance but also a low AP processing latency.

This paper proposes a cell-free parallel scheme for radio stripes. In this scheme, AP computes major parts of signal processing in parallel and only requires to serially add the signal from the previous stage. MR processing can be easily implemented in this scheme, but its performance is bad. To gain a better performance, this paper proposes two efficient parallel schemes, which are respectively based on MR and LMMSE. The first one is interference-aware MR (I-MR) processing with an enhancement of large-scale prior knowledge. For this scheme, a UC strategy using large-scale signal-to-interference-and-noise ratio (SINR) is also proposed. The second is distributed regularized zero-forcing (D-RZF) scheme where AP transmits not only MR data streams but also a inter-user interference matrix. At CPU, a RZF processing deriving from LMMSE is done with these two signals from APs.

The contribution of this paper is as follows: (a) The concept of parallel scheme for radio stripes is first proposed. It helps to greatly reduce the processing latency via efficient parallel processing at all APs; (2) I-MR is proposed to gain a better performance than MR. Even using much less APs, the proposed I-MR scheme gains a better performance than that of MR processing; (3) D-RZF is proposed via approximating LMMSE. The proposed D-RZF scheme first gains a performance very close to that of LMMSE with a low front-haul loading close to that of MR processing. The rest of this paper is organized as follows. In Section II, system model and existing works are introduced. In Section III, I-MR and D-RZF are proposed and analyzed. In Section IV, the simulation results are shown to verify the performance and complexity. In Section V, this paper is briefly concluded. In this paper, $(\cdot)^*$, $(\cdot)^T$, $(\cdot)^H$, and $\|\cdot\|$ denote conjugate, transpose, Hermitian transpose, and $l_2$ norm of a matrix or vector, respectively. $\mathbf{I}_N$ is the identity matrix of size $N \times N$, and $\mathbb{E}\{\cdot\}$ denote the expectation function.

## II. SYSTEM MODEL AND EXISTING WORKS

### A. System Model

As shown in Fig. 1(a), a radio stripe is deployed in a large room, and there are $L$ APs and $K$ active single-antenna UEs. Each AP has $N$ antennas, and the total antenna number of all APs is $M = NL$. The channel vector of user $k$ is $\mathbf{h}_k = [\mathbf{h}_{k,1}^T, \mathbf{h}_{k,2}^T, ..., \mathbf{h}_{k,L}^T]^T \in \mathbb{C}^{M \times 1}$, where $\mathbf{h}_{k,l} \in \mathbb{C}^{N \times 1}$ is the channel between user $k$ and AP $l$. A Rayleigh block fading channel of $\mathbf{h}_{k,l} \sim \mathcal{CN}(\mathbf{0}, \mathbf{R}_{k,l})$ is assumed, where $\mathbf{R}_{k,l} \in \mathbb{C}^{N \times N}$ is the spatial covariance matrix. The large-scale coefficient $\beta_{k,l} = \text{tr}(\mathbf{R}_{k,l})/N$. When $N$ antennas at each AP are uncorrelated, $\mathbf{R}_{k,l} = \beta_{k,l}\mathbf{I}_N$; when $N$ antennas are correlated, $\mathbf{R}_{k,l}$ can be obtained using the correlated scattering models in [8].

The channel is assumed to be flat over a coherence block of $\tau_c$ channel uses. Among $\tau_c$ channel uses, $\tau_p$ of them are used for pilot, while $(\tau_c - \tau_p)$ are for payload data. In this paper, pilot reuse is not considered, which means $K \leq \tau_p$. Pilot reuse can be supported via using a prior knowledge of large-scale channel correlation to estimate channels [7]. However, to gain a good performance, an extra pilot allocation strategy minimizing the pilot contamination impact is required, which is out of the scope of this paper. An orthogonal pilot set of $\{\phi_1, \phi_2, ..., \phi_{\tau_p}\}$ with $\phi_k \in \mathbb{C}^{\tau_p}$ and $\|\phi_k\|^2 = \tau_p$. User $k$ is assumed to use pilot $\phi_k$ with $k \leq \tau_p$.

The received signal of pilot $\mathbf{Z}_l \in \mathbb{C}^{N \times \tau_p}$ at AP $l$ is

$$\mathbf{Z}_l = \sum_{i=1}^{K} \sqrt{p_i} \mathbf{h}_{i,l} \phi_i^T + \mathbf{N}_l \quad (1)$$

where $p_i \geq 0$ is the transmit power, and $\mathbf{N}_l \in \mathbb{C}^{N \times \tau_p}$ is the complex Gaussian noise with independent entries following a distribution of $\mathcal{CN}(\mathbf{0}, \sigma^2)$.

Similarly, the received signal of data symbols $\mathbf{y}_l \in \mathbb{C}^N$ at AP $l$ is

$$\mathbf{y}_l = \sum_{i=1}^{K} \sqrt{p_i} \mathbf{h}_{i,l} s_i + \mathbf{n}_l \quad (2)$$

where $s_i \in \mathbb{C}$ is the modulated symbols from user $i$ with $\mathbb{E}\{|s_i|^2\} = 1$, and $\mathbf{n}_l \in \mathbb{C}^N$ is the complex Gaussian noise with independent entries distributed as $\mathcal{CN}(\mathbf{0}, \sigma^2)$.

### B. Existing Works

The MMSE channel estimation [8] $\hat{\mathbf{h}}_{k,l} \in \mathbb{C}^{N \times 1}$ between user $k$ and AP $l$ is

$$\hat{\mathbf{h}}_{k,l} = \sqrt{p_k} \mathbf{R}_{k,l} \left(\tau_p p_k \mathbf{R}_{k,l} + \sigma^2 \mathbf{I}_N\right)^{-1} \mathbf{Z}_l \phi_k^* \quad (3)$$

With the channel estimation of each AP, the combination vector $\mathbf{v}_k$ for user $k$ can be obtained to combine the all received signals $\mathbf{y} = [\mathbf{y}_1^T, \mathbf{y}_2^T, ..., \mathbf{y}_L^T]^T \in \mathbb{C}^{M \times 1}$, and the recovered data of user $k$ is $\mathbf{v}_k^H \mathbf{y}$.

Assume $\hat{\mathbf{h}}_k = [\hat{\mathbf{h}}_{k,1}^T, \hat{\mathbf{h}}_{k,2}^T, ..., \hat{\mathbf{h}}_{k,L}^T]^T \in \mathbb{C}^{M \times 1}$. The combination vector of user $k$ using MR processing [2] is $\mathbf{v}_k^{MR} = \hat{\mathbf{h}}_k$, while that using optimal LMMSE processing [4] is

$$\mathbf{v}_k^{MMSE} = \left(\sum_{i \neq k} p_i \left(\hat{\mathbf{h}}_i \hat{\mathbf{h}}_i^H + \mathbf{R}_i - \mathbf{\Gamma}_i\right) + \sigma^2 \mathbf{I}_M\right)^{-1} \hat{\mathbf{h}}_k \quad (4)$$

where $(\mathbf{R}_i - \mathbf{\Gamma}_i)$ is the channel correlation estimation error matrix. $\mathbf{R}_i \in \mathbb{C}^{M \times M}$ and $\mathbf{\Gamma}_i \in \mathbb{C}^{M \times M}$ are

$$\mathbf{R}_i = \begin{bmatrix} \mathbf{R}_{i,1} & \mathbf{0} & ... & \mathbf{0} \\ \mathbf{0} & \mathbf{R}_{i,2} & ... & \mathbf{0} \\ ... & ... & ... & ... \\ \mathbf{0} & \mathbf{0} & ... & \mathbf{R}_{i,L} \end{bmatrix} \quad (5)$$

$$\mathbf{\Gamma}_i = \tau_p p_i \mathbf{R}_i \left(\tau_p p_i \mathbf{R}_i + \sigma^2 \mathbf{I}_M\right)^{-1} \mathbf{R}_i \quad (6)$$

The front-haul loading of MR processing is $K(\tau_c - \tau_p)$, while that of centralized LMMSE processing is $M(\tau_c - \tau_p) + MK$. As $K < M$ in a typical cell-free system, e.g. $M = 96$, $K = 10$ in [7], MR processing is able to reduce the front-haul. However, MR performs much worse than LMMSE [4].

To gain both good performance and low front-haul loading, [7] proposes N-LMMSE. The main idea is to compute LMMSE combination of $N$ streams for $K$ users in AP 1, and normalize the noise power of combined streams to $\sigma^2$. Apart from the data estimations of $K$ users, the estimated effective channels and effective channel errors variances are required to transmit to the next stage. For $l \geq 2$, AP $l$ does a similar operations to $(N+1)$ received steams including $N$ from received antennas and one from AP $l$-1. The operations of N-LMMSE are relatively complex and are not shown here. It can be found in [7] if interested. Using N-LMMSE, a performance between those of MR and LMMSE can be obtained, while the front-haul loading is $K(\tau_c - \tau_p) + 2K^2$, which is close to that of MR when $K << (\tau_c - \tau_p)$.

## III. EFFICIENT PARALLEL SCHEMES

### A. Parallel Schemes

Although N-LMMSE is able to gain a good performance with a relatively low front-haul loading, most signal processing operations require to be done serially. This sequential processing leads to a large processing latency from AP 1 to CPU. When the AP number $L$ is large, and AP hardware is simple, the AP processing latency becomes a severe problem.

To solve this problem, a parallel scheme is proposed for radio stripes as shown in Fig. 1(b). In this scheme, major

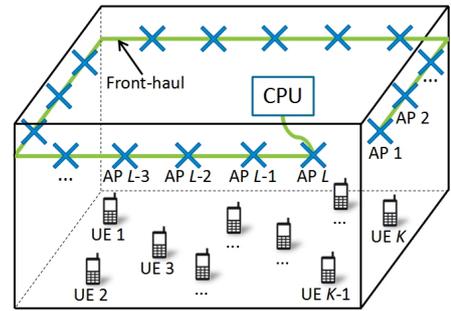

(a) Radio stripe implementation in a large room

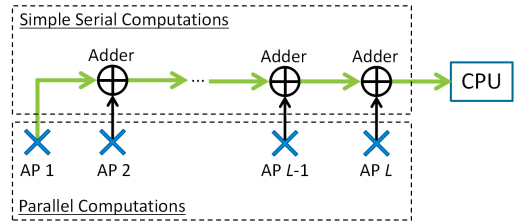

(b) The proposed parallel scheme

Fig. 1 Cell-free massive MIMO using radio stripe implementations. UE represents user equipment.

part of of AP computations are done in parallel. Only simple adding operations are serial, which will not lead to a large latency. Obviously, MR can be implemented in this scheme as

$$\hat{s}_k^{\text{MR}} = \hat{\mathbf{h}}_k^H \mathbf{y} = \sum_{i=1}^{L} \hat{\mathbf{h}}_{k,i}^H \mathbf{y}_i \quad (7)$$

However, the performance of MR is not good. Therefore, better methods for this scheme are required, which are named efficient parallel schemes.

*B. Interference-aware MR*

The first proposed efficient parallel scheme is based on MR processing. Conventional MR processing only maximizes the receiving SNR without considering interference. To improve the performance, interference should be considered, and I-MR is proposed using the prior knowledge large-scale fading coefficient. The large-scale information varies very slow, and it can be obtained in advance. The combination vector of the proposed I-MR is

$$\mathbf{v}_k^{\text{IMR}} = \left[\frac{\hat{\mathbf{h}}_{k,1}^T}{\sum_{i \neq k} \beta_{i,1} + \sigma^2}, \frac{\hat{\mathbf{h}}_{k,2}^T}{\sum_{i \neq k} \beta_{i,2} + \sigma^2}, \ldots, \frac{\hat{\mathbf{h}}_{k,L}^T}{\sum_{i \neq k} \beta_{i,L} + \sigma^2}\right]^T \quad (8)$$

The weights are smaller for the received streams having larger prior multi-user interference. Using (8), the low front-haul loading of MR is still kept.

When I-MR is used, a novel UC strategy of sorting prior SINR is employed. The activated AP ratio for one user is assumed to be $\alpha$. The activated AP index set $\mathbf{\Omega}_k$ consisting of $\alpha L$ APs for user $k$ can be selected by CPU as

$$\mathbf{\Omega}_k = \arg\max_{1 \leq l \leq L}^{\alpha L} \frac{\beta_{k,l}}{\sum_{i \neq k} \beta_{i,l} + \sigma^2} \quad (9)$$

Then, CPU informs each AP to process part of users. UC strategy is designed to reduce the front-haul loading of cell-free system, while it cannot reduce that of the proposed parallel scheme where the front-haul loading is decided by $K$, instead of $M$. It still can be used to reduce the computational complexity of AP in parallel scheme. That is to say, UC strategy reduces the power consumption of signal processing for APs, which is also very useful in practice.

*C. Distributed RZF*

The second proposed efficient parallel scheme is derived from LMMSE processing. Conventional LMMSE requires centralized signal processing, which leads to a high front-haul overloading for radio stripes. This paper aims at finding a way to gain LMMSE performance with limited front-haul loading and parallel scheme. First, consider the scenario of uncorrelated antennas at each AP. That is to say, $\mathbf{R}_{k,l} = \beta_{k,l}\mathbf{I}_N$, and (4) becomes

$$\mathbf{v}_{\text{IID}-k}^{\text{MMSE}} = \left(\sum_{i \neq k} \left(p_i \hat{\mathbf{h}}_i \hat{\mathbf{h}}_i^H + \begin{bmatrix} \varphi_{i,1}\mathbf{I}_N & & & \\ & \varphi_{i,2}\mathbf{I}_N & & \\ & & \ddots & \\ & & & \varphi_{i,L}\mathbf{I}_N \end{bmatrix}\right) + \sigma^2 \mathbf{I}_M \right)^{-1} \hat{\mathbf{h}}_k \quad (10)$$

where $\varphi_{i,l} = \frac{\beta_{i,l}\sigma^2}{\tau_p p_i \beta_{i,l} + \sigma^2}$.

Usually, $\tau_p p_i \beta_{i,l}/\sigma^2 \gg 1$, e.g. it is 20 when $\tau_p = 20$ and $p_i \beta_{i,l}/\sigma^2 = \text{SNR} = 0$ dB. Then, an approximation is made as

$$\varphi_{i,l} \approx \frac{\beta_{i,l}\sigma^2}{\tau_p p_i \beta_{i,l}} = \frac{\sigma^2}{\tau_p p_i} \quad (11)$$

For same $i$ and different $l$, $\varphi_{i,l}$ is a similar. Then, $\varphi_{i,l}$ is replaced by an average value, and an approximation of (11) is made as

$$\tilde{\mathbf{v}}_{\text{IID}-k}^{\text{MMSE}} = \left(\sum_{i \neq k} \left(p_i \hat{\mathbf{h}}_i \hat{\mathbf{h}}_i^H + \begin{bmatrix} \overline{\varphi}_i \mathbf{I}_N & & & \\ & \overline{\varphi}_i \mathbf{I}_N & & \\ & & \ddots & \\ & & & \overline{\varphi}_i \mathbf{I}_N \end{bmatrix}\right) + \sigma^2 \mathbf{I}_M \right)^{-1} \hat{\mathbf{h}}_k$$

$$= \left(\sum_{i \neq k} p_i \hat{\mathbf{h}}_i \hat{\mathbf{h}}_i^H + \left(\sum_{i \neq k} \overline{\varphi}_i + \sigma^2\right) \mathbf{I}_M \right)^{-1} \hat{\mathbf{h}}_k \quad (12)$$

where $\overline{\varphi}_i = \frac{1}{L}\sum_{l=1}^{L} \varphi_{i,l}$. (12) can also be seen as a regularized zero-forcing (RZF) receiver, and the derivation step shows that it can gain a performance close to LMMSE.

Using Sherman-Morrison formula [9], a modified version of (12) can be easily obtained as

$$\hat{\mathbf{v}}_{\text{IID}-k}^{\text{MMSE}} = \left(\sum_{i=1}^{K} p_i \hat{\mathbf{h}}_i \hat{\mathbf{h}}_i^H + \left(\sum_{i \neq k} \overline{\varphi}_i + \sigma^2\right) \mathbf{I}_M \right)^{-1} \hat{\mathbf{h}}_k = \eta \tilde{\mathbf{v}}_{\text{IID}-k}^{\text{MMSE}} \quad (13)$$

where $\eta \in \mathbb{C}$ is a coefficient, which does not affect the performance.

Assume $\hat{\mathbf{V}}_{\text{IID}}^{\text{MMSE}} = \left[\hat{\mathbf{v}}_{\text{IID}-1}^{\text{MMSE}^T}, \ldots, \hat{\mathbf{v}}_{\text{IID}-K}^{\text{MMSE}^T}\right]^T \in \mathbb{C}^{M \times K}$, and (13) can be written in another form [8] as

$$\hat{\mathbf{V}}_{\text{IID}}^{\text{MMSE}} = \hat{\mathbf{H}} \left(\hat{\mathbf{H}}^H \hat{\mathbf{H}} + \left(\sum_{i \neq k} \overline{\varphi}_i + \sigma^2\right) \mathbf{I}_K \right)^{-1} \quad (14)$$

where $\hat{\mathbf{H}} = \left[\sqrt{p_1}\hat{\mathbf{h}}_1, \sqrt{p_2}\hat{\mathbf{h}}_2, \ldots, \sqrt{p_K}\hat{\mathbf{h}}_K\right] \in \mathbb{C}^{M \times K}$.

Assume $\hat{\mathbf{s}}_{\text{IID}}^{\text{MMSE}} = \left[\hat{s}_{\text{IID}-1}^{\text{MMSE}}, \hat{s}_{\text{IID}-2}^{\text{MMSE}}, \ldots, \hat{s}_{\text{IID}-K}^{\text{MMSE}}\right]^T \in \mathbb{C}^{K \times 1}$. The data symbol estimation vector of all users is

$$\hat{\mathbf{s}}_{\text{IID}}^{\text{MMSE}} = \left(\hat{\mathbf{V}}_{\text{IID}}^{\text{MMSE}}\right)^H \mathbf{y} = \left(\hat{\mathbf{H}}^H \hat{\mathbf{H}} + \left(\sum_{i \neq k} \overline{\varphi}_i + \sigma^2\right) \mathbf{I}_K \right)^{-1} \hat{\mathbf{H}}^H \mathbf{y}$$

$$= \left(\hat{\mathbf{H}}^H \hat{\mathbf{H}} + \left(\sum_{i \neq k} \overline{\varphi}_i + \sigma^2\right) \mathbf{I}_K \right)^{-1} \hat{\mathbf{s}}^{\text{MRC}} \quad (15)$$

where $\hat{\mathbf{s}}^{\text{MRC}} = \left[\hat{s}_1^{\text{MRC}}, \hat{s}_2^{\text{MRC}}, \ldots, \hat{s}_K^{\text{MRC}}\right]^T \in \mathbb{C}^{K \times 1}$. Using the prior knowledge of large-scale channel, $\left(\sum_{i \neq k} \overline{\varphi}_i + \sigma^2\right) \mathbf{I}_K$ is obtained which can be calculated in advance at CPU. $\hat{s}_k^{\text{MRC}}$ can be calculated in parallel as shown in (7). The remaining item in (15) can also be obtained in parallel as

$$\hat{\mathbf{H}}^H \hat{\mathbf{H}} = \sum_{i=1}^{L} \hat{\mathbf{H}}_i^H \hat{\mathbf{H}}_i \quad (16)$$

where $\hat{\mathbf{H}}_i = \left[\sqrt{p_1}\hat{\mathbf{h}}_{1,i}, \sqrt{p_2}\hat{\mathbf{h}}_{2,i}, \ldots, \sqrt{p_K}\hat{\mathbf{h}}_{K,i}\right] \in \mathbb{C}^{N \times K}$. Thus, (15) can be calculated using proposed parallel scheme. After obtaining the all three items in (15), the remaining calculation is done by CPU. This scheme is named D-RZF. It can be implemented using parallel scheme of Fig. 1(b), as (7) and (16) are simultaneously done in different APs and adds up during front-haul transmissions.

For APs using correlated antennas, the $M$-dimension channel correlation error matrix ($\mathbf{R}_i - \mathbf{\Gamma}_i$) is not diagonal, which makes it difficult to be converted into the form of (10).

As this channel correlation error matrix should be much smaller than the channel correlation estimation matrix $\hat{\mathbf{h}}_i\hat{\mathbf{h}}_i^H$, $(\mathbf{R}_i - \mathbf{\Gamma}_i)$ is roughly approximated by its diagonal matrix. That is to say, (10) can still be used, and (15) also works for correlated antennas at AP. The simulation results in the next section show it gains a throughput very close to centralized optimal LMMSE of (4) no matter whether antennas at AP are correlated or uncorrelated.

The front-haul loading of D-RZF is $K(\tau_c - \tau_p) + K^2$, while that of I-MR is $K(\tau_c - \tau_p)$. D-RZF performs better than I-MR, and they two can be used in different scenarios. When $K$ is much smaller than $(\tau_c - \tau_p)$, D-RZF uses a front-haul loading similar to that of MR/I-MR and thus more preferred for better performance; when $K$ is comparable to $(\tau_c - \tau_p)$ and the performance requirement is not strict, I-MR can be used for a lower front-haul loading in this case.

## IV. SIMULATION RESULTS

### A. Simulation Settings

The simulation scenario is shown in Fig. 1(a). The radio stripe is placed around the top edges of surrounding walls. The radio stripe length is 500m, and the room size is 125m×125m×5m in the order of length, width and height. $L = 24$, $N = 4$, $M = 96$, $K = 20$ or $40$, $\tau_c = 200$, and $\tau_p = K$. APs are uniformly distributed at the radio stripe, while users are randomly distributed inside the floor of the room. The correlated channel is generated by Gaussian correlated scattering model [7] with a angle spread of 15°. The large-scale fading coefficient (dB) $\beta_{k,l} = -30.5 - 36.7\log_{10}(d_{k,l})$, where $d_{k,l}$ is the distance (m) between user $k$ and AP $l$. The transmit power of each user is 200 mW, and the noise power $\sigma^2 = -92$ dBm. Both correlated and uncorrelated antennas are considered at APs to verify the performance of proposed schemes. When UC or novel UC is used, the percentage of activated APs $\alpha$ is set to 1/4. The spectral efficiency (SE) of user $k$ is calculated via

$$SE_k = \frac{\tau_c - \tau_p}{\tau_c} E\{\log_2(1 + SINR_k)\} \quad (17)$$

where $SINR_k$ is the receiving SINR of user $k$. The complexity is represented by the number of complex number multiplications, while the total time complexity of all APs is calculated by

$$C_T = L \cdot C_{Serial} + C_{Parallel} \quad (18)$$

where $C_{Serial}$ and $C_{Parallel}$ denote the complexity of calculations in each AP that require to be done serially and can be done in parallel, respectively.

### B. Performance Comparisons

Fig. 2 shows the cumulative distribution of single user SE for different cell-free schemes. In uncorrelated channels scenario of $K = 20$, MR plus UC performs worse than MR as the activated APs number is reduced to $L/4$. As a comparison, the proposed I-MR plus novel UC is able to gain a better performance than MR, even when the number of APs per user is reduced by 75%. Also, the proposed D-RZF is able to perform as good as the centralized LMMSE, which is much better than that of N-LMMSE. When UC strategy is used, N-LMMSE gains a performance similar to that of MR, while the proposed D-RZF also performs much better than that.

In correlated channels scenario, the performance of all schemes degrade due to channel correlation, while similar observations can also be found. D-RZF still gains the performance of centralized LMMSE, although channel correlation prior knowledge is omitted in (15). It shows the impact of channel correlation on the combination vector calculation is tiny in this cell-free massive MIMO using radio stripes. It means that the rough approximation of only the diagonal matrix for $(\mathbf{R}_i - \mathbf{\Gamma}_i)$ does not affect the performance. The performance of N-LMMSE plus UC becomes better than that of MR in this case, but it is still much worse than D-RZF plus UC. The results from Fig. 2 show that both the proposed two efficient parallel schemes perform better than their counterparts.

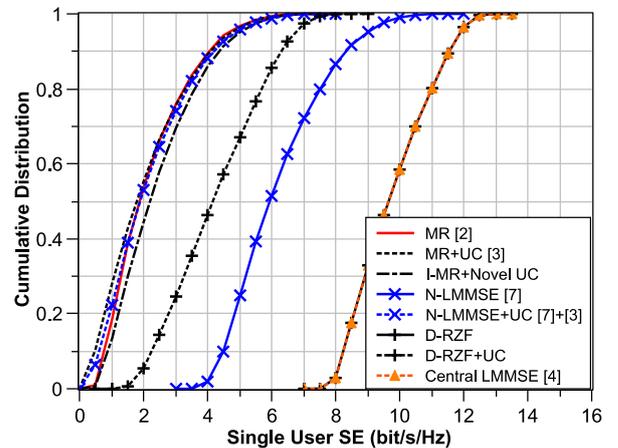

(a) Uncorrelated Channels

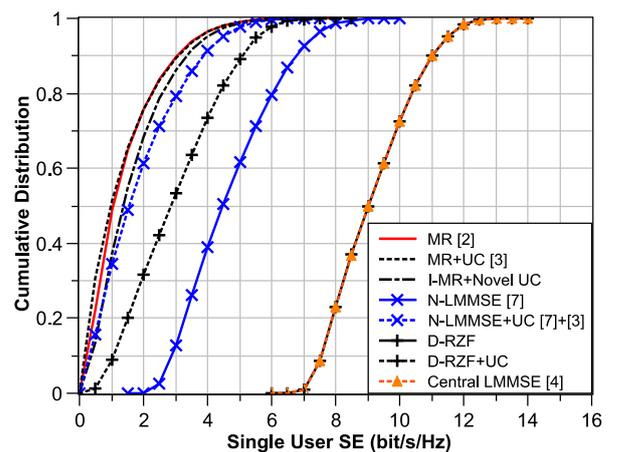

(b) Correlated Channels

Fig. 2 Single user spectral efficiency comparison of different cell-free schemes in the form of cumulative distribution for $K = 20$.

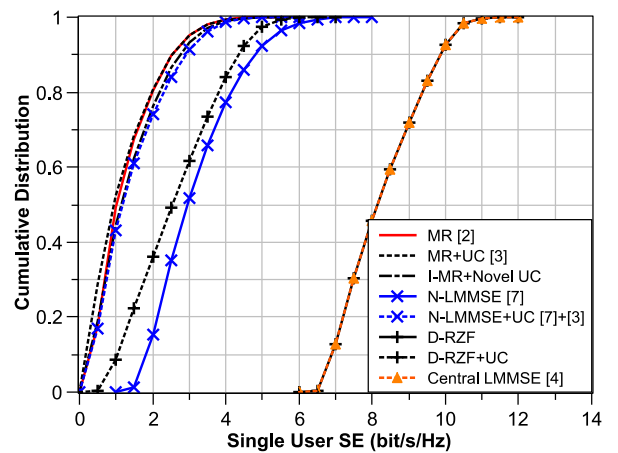

Fig. 3 Single user spectral efficiency comparison of different cell-free schemes in the form of cumulative distribution for $K = 40$ and uncorrelated channels.

In Fig. 3, the simulation result of $K = 40$ and uncorrelated channels is also shown. Compared with that of $K = 20$, the general trend remains the same. One major difference is that the performance gap between N-LMMSE and D-RZF gets larger, and the performance of D-RZF plus UC becomes closer to that of N-LMMSE without UC. When $K$ changes from 20 to 40, the average single user SE of N-LMMSE reduces from 9.78 to 8.28 bits/s/Hz, which is around 15.3% SE degradation. Note that the payload length decreases from 180 to 160, which bring about 11.1% SE reduction. That is to say, the channel qualities of different users decrease only about 4.2% when $K$ increases from 20 to 40. As a comparison, the average single user SE changes from 6.19 bits/s/Hz to 3.16 bits/s/Hz, which is an almost 50% SE degradation. This result means that the performance advantage of D-RZF over N-LMMSE becomes larger when the number of users $K$ increases.

## C. Cost and Complexity Comparison

To implement cell-free massive MIMO using radio stripes in practice, cost and power consumption are two crucial factors to consider. Front-haul cost is decided by its loading, while the cost and power consumption of AP are decided by AP complexity. As mentioned before, the processing time of many low-cost APs could add up to a large delay in radio stripes, which also means more expensive AP hardware is required when transmission latency is restricted. These aspects are evaluated via three indicators including (i) front-haul loading, (ii) computational complexity per AP and (iii) total time complexity of all APs. The front-haul loading is normalized by the total number of receiving data symbols, which is $K \cdot (\tau_c - \tau_p)$.

The front-haul loading comparison is shown in Fig. 4(a). MR, as well as I-MR, is able to gain a minimum front-haul loading of 1. As a comparison, centralized LMMSE requires a loading of $M/K$ to transmit one data symbol, which makes it not suitable for radio stripes. Both N-LMMSE and D-RZF are able to greatly reduce the front-haul loading especially when $\tau_c$ is large. For $\tau_c = 200$, the front-haul loadings of N-LMMSE and D-RZF are only 1.11 and 1.22, respectively. Compared with N-LMMSE, the proposed D-RZF is able to gain a even lower front-haul loading.

The complexity comparison is shown in Fig. 4(b) and Fig. 4(c). When UC is not used, MR and I-MR gain the lowest computational complexity, and that of D-RZF is very close to it. N-LMMSE has more than two-fold computational complexity than other schemes as most APs require to compute LMMSE of $(N+1)$ data streams for every user, which include the computation of matrix inverse. When UC or Novel UC strategy is applied, all schemes reduce the average computation complexity by about 75%. This comparison shows that the complexity in one AP of D-RZF is still very close to that of MR.

The total time complexity is important for latency using same AP hardware, or the AP hardware cost for same latency requirement. As mentioned before, MR, I-MR and D-RZF can be implemented in parallel schemes, and the total time complexity is the same as complexity per AP according to (18). As a comparison, N-LMMSE has at least

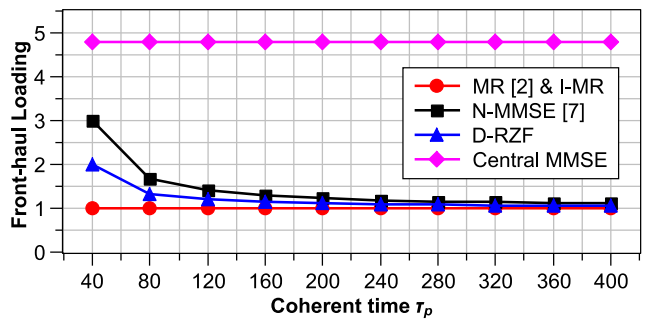
(a) Front-haul loading per receiving data symbol

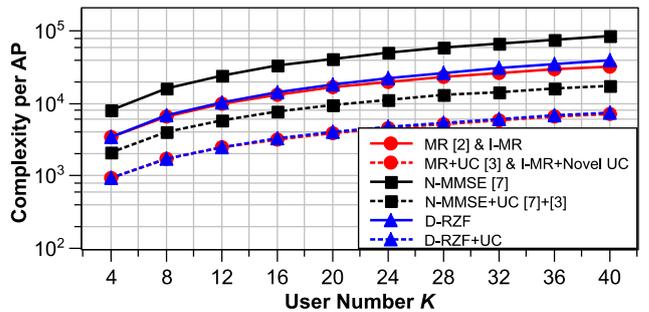
(b) Computational complexity per AP

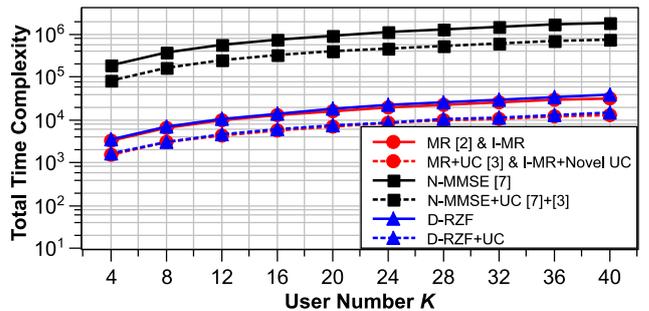
(c) Total time complexity of all APs

Fig. 4 The cost and complexity comparison of different schemes. The common parameters are listed in Section II-A. In (a), $K = 20$.

40 times more total time complexity than other schemes. The reason is that only the channel estimation operation can be done in parallel, while other computations requires a serial processing. The total time complexities are not reduced to their 1/4 when UC strategies are used. Each AP has different numbers of users to process, and latency is decided by the one which has the most number. This number is obtained from simulations, which shows that UC still can reduce the total time complexities of different schemes by around a half for different schemes.

## V. CONCLUSIONS

Cell-free massive MIMO using radio stripes is a very promising B5G/6G technology. Key factors of cost and complexity are crucial for practical implementations. This paper proposes a parallel scheme for cost and complexity. To ensure the performance, two efficient signal processing schemes are proposed. They are respectively based on MR and LMMSE, and can be used in different scenarios. They are also verified via simulations in terms of both performance and complexity. Compared with sequential processing for radio stripes, the proposed two parallel schemes can reduce front-haul loading, complexity and total processing latency. Moreover, the proposed D-RZF is able to perform much better than its counterparts, as well as gains a performance similar to LMMSE.